# Effect of Nanoscale Confinement on the $\beta$-$\alpha$ Phase Transition in Ag$_2$Se

*Vincent Leon* (1), *Yang Ren* (2) *and Marie-Louise Saboungi* (1)

* Corresponding author: saboungi@cnrs-orleans.fr

(1) Centre de Recherche sur la Matière Divisée, CNRS-Université d'Orléans, 1B rue de la Férollerie, 45071 Orléans Cedex 2, France

(2) Advanced Photon Source, Argonne National Laboratory, 9700 S. Cass Ave., Argonne, IL 60439, USA

**Abstract**

The confinement of silver selenide was investigated using mesoporous silica. Results from x-ray diffraction and electron microscopy show that the confined material still exhibits a $\beta$ to $\alpha$ transition similar to the one that takes place in the bulk crystalline state but with a transition temperature that depends significantly on the confinement conditions. Decreasing the pore size leads to an increase of the transition temperature, opposite to the behavior of the melting point observed in several metallic and organic materials. In the free particles, on the other hand, no size dependence is observed with particle sizes down to 4 nm.

A recent review by Alcoutlabi and McKenna[1] has summarized the state of the art of nanoscale confinement effects on the first-order thermodynamics in confined geometries and shown that the Gibbs-Thomson equation is suitable for describing the changes in melting phenomena observed in the literature due to nanometer scale confinement. They also discussed the more challenging issue of the glass transition for confined systems under different geometrical conditions. However, to our knowledge the problem of the changes in solid-solid phase transitions due to confinement has not been addressed. Some silver ionic salts as well as silver chalcogenides undergo a relatively low-temperature solid-solid transition with dramatic differences between the two phases, e.g., transition from an ordered to a disordered phase with changes in the nature of the conduction mechanism in each phase.[2,3] The structure and spectroscopic properties of various Se based compounds confined in cages of zeolites have been studied,[4] but no attempts have been made to investigate changes induced by the confinement on first-order transitions. Silver selenide, Ag$_2$Se, is such an example and here we focus uniquely on this compound. At room temperature, Ag$_2$Se is a narrow band-gap semiconductor with high carrier mobility and without magnetoresistance (MR) and undergoes a phase transition in the solid state.[5,6] The low-temperature phase, designated as $\beta$–Ag$_2$Se with an orthorhombic structure, has been widely used as a photosensitizer in photographic films and thermochromic materials. The transition to the high-temperature phase $\alpha$–Ag$_2$Se takes place accompanied by a decrease in volume and has been reported with values of the transition temperature T$_{\beta\alpha}$ ranging from 127 °C to 143 °C.[7,8] The high-temperature phase is a superionic conductor with mobile Ag$^+$ ions used as the solid electrolyte in photochargeable secondary batteries.[9] In the liquid state, it has been shown that the temperature dependence of the conductivity is negative, which is unusual for liquids with a conductivity in the range typical for liquid semiconductors in the narrow definition of Mott.[10,11] Moreover, $\beta$–Ag$_2$Se has been recently shown to exhibit a significant magnetoresistance (MR) by manipulating small departures from stoichiometry.[12] This large magnetic response, a near-linear increase of the resistance with the applied magnetic field without saturation at high fields, is comparable in magnitude to the colossal MR observed in manganese perovskites,[13] but occurring here in a non-magnetic material.[14]

Here we report a procedure for confining $Ag_2Se$ in an inorganic matrix and a study of the effects of confinement on the phase transition temperature. Porous silica was chosen as the matrix because it is chemically inert, the size and shape of the pores are well controlled [15] and, in the case of selenium-based particles, the interactions between silica and selenium are very weak.[16] In order to achieve confinement, we performed a direct synthesis of silver chalcogenide inside the pores of a mesoporous silica presenting an ordered array of pores with a diameter between 2 and 10 nm. This procedure has the advantage of taking place at room temperature under soft synthesis conditions. Mesoporous silica (SBA-15) was prepared following the procedure described by Zhao et al.[17] The choice of SBA-15 as porous silica host instead of, e.g., MCM-41 was dictated by the reported higher stability of the walls and the presence of microporous connections between the mesopores.[18] Typically, 3 g of porosity agent are dissolved in 22.5 g of water and diluted in 90 g of HCl (pH 2). The porosity agents used here are poly(ethylene oxide)-poly(propylene oxide)-poly(ethylene oxide) triblock copolymers with different chain lengths to provide different pore sizes.[19] Then 6.375 g of the silica precursor TEOS (tetraethoxysilane, 99.99%, Alfa Aesar) are added under vigorous stirring for 20 h at 35 °C. The solution is aged under different conditions, summarized in Table 1, depending on the pore size needed. Finally, the obtained gel is calcinated at 550 °C for 12 hours to remove the polymer template. For each sample, surface areas were calculated using the BET adsorption-desorption method and pore size distributions were analyzed by the BJH method. An impregnation technique to obtain some polyvinyl alcohol–selenide nanocomposites, was used to synthesize a composite material with silver selenide inside silica pores. [20]Silver nitrate ($AgNO_3$) and sodium selenosulfate ($Na_2SeSO_3$) were used as silver and selenium sources respectively. Typically, 500 mg of mesoporous silica is placed in a 50 ml flask and 5 ml of an aqueous solution of $AgNO_3$ (0.1 M) is added, under constant stirring at room temperature for 3 h. Then 0.5 ml of an aqueous solution of $Na_2SeSO_3$ (pH 10) is added under the same conditions (room temperature, 3 h stirring). This solution is prepared by adding 4 g (0.05 mol) of selenium powder to 50 ml of a 1M aqueous solution of $Na_2SO_3$. The mixture is refluxed at 70 °C for 24 h and then filtrated under vacuum and stored at 60 °C to avoid decomposition. A few drops of ammonia solution are added to reach a pH value of 10. The dark solution of sodium selenosulfate is then filtrated, washed with deionized water to remove solvents, and dried under vacuum. $Ag_2Se$ was synthesized inside pores of different size, but the conditions just described were the same for each sample.

In order to determine whether the $Ag_2Se$ is inserted into the silica pores, we used transmission electron microscopy (Philips CM20) and scanning electron microscopy (SEM, Hitachi). Figure 1 clearly shows $Ag_2Se$ inside 2 nm-size mesopores; the fact that some parts of the pores are not filled with $Ag_2Se$ just indicates that the proportions of silica and aqueous solutions of silver and selenium ions have to be adjusted to have a homogenous filling of the pores. It can also be seen that the material fills the pores as nanowires rather than nanoparticles. Consequently, it is interesting to compare the results obtained for these samples with both bulk $Ag_2Se$ and $Ag_2Se$ nanoparticles. For the synthesis of $Ag_2Se$, we used the method described by Dusastre et al.,[21] where silver and selenium powders are mixed in stoichiometric proportion in *n*-butylamine as solvent (or more precisely surface activator) to give particles with an average size of about 150 nm (Figure 2). We synthesized $Ag_2Se$ nanoparticles using the double-microemulsion method[22,23] involving microemulsions of $Ag^+$ and $Se^{2-}$ in a ternary AOT/*n*-heptane/water system. This method is suitable for controlling the size of the particles by adjusting the ratio of surfactant (AOT) to water. Here, we chose to generate nanoparticles with a reasonably narrow size distribution around a diameter of 3.9 nm (Figure 3).

To study the variations in $T_{\beta\alpha}$ for the different conditions of $Ag_2Se$, we made structural measurements using the high-energy X-ray diffractometer at beam line 11-ID-C of the Advanced Photon Source at a photon energy of 115 keV. Powdered samples were placed in thin capillaries (1 to 3 mm diameter) and x-ray diffraction (XRD) measurements were carried out at different temperatures using a controlled heater. During a typical measurement, the temperature is increased by 1 K in the region the phase transition temperature, up to 200°C, and then cooled down to room temperature with the same protocol. For the bulk, XRD measurements show that $T_{\beta\alpha}$ takes place at around 140 ±1°C. As shown in Figure 4, the transition between $\beta$-Ag$_2$Se and $\alpha$–Ag$_2$Se for 150 nm-sized particles occurs also at around 140 °C : the three main peaks characterizing $\beta$– Ag$_2$Se at $Q$ = 23.9 (112), 27.9 (031) and 30.0 nm$^{-1}$ (113) are replaced by the two main peaks of $\alpha$–Ag$_2$Se at $Q$ = 25.3 (200) and 30.9 nm$^{-1}$ (11). It appears that the size effect on $T_{\beta\alpha}$ is negligible for free particles. This lack of size dependence is not universal since, in the case of an oxide like SrBi$_2$Ta$_2$O$_9$, for example, the ferroelectric phase transition temperature is strongly dependent on the diameter of the particles.[24]

Using the same protocol, XRD measurements for silver selenide confined inside 8, 3.5 and 2 nm diameter pores were carried out showing that $T_{\beta\alpha}$ is 141, 142 and 146 °C, respectively. For illustration purpose, Figure 5 shows a typical spectra collected in the case of 2 nm pores and Figure 6 the dependence of $T_{\beta\alpha}$ on pore size. The measured increase of $T_{\beta\alpha}$ with decreasing pore size contrasts with the lack of size effect found for the free particles. A size effect on the phase transition of some metallic particles have been reported, with a decrease in melting point with decreasing particle size in gold, for example.[25] It has also been shown that the melting transition of some confined organic molecules has the opposite behavior than the one found here.[26,27] Results for the effect of confinement on the glass transition appear to be more complicated, with different behaviors observed for the same material depending on the experimental methods used.[1]

In conclusion, we have developed a procedure for direct synthesis of Ag$_2$Se inside the pores of a mesoporous silica containing an ordered array of pores with a diameter between 2 and 10 nm, and found a steady decrease in the temperature of the β→α transition with decreasing pore size. On the other hand, no such effect is observed in free nanoparticles down to a size of 3.9 nm. Since the β→α transition can be regarded in some sense as a melting of the Ag$^+$ sublattice, it is interesting that the behavior of both types of nanoscale Ag$_2$Se is quite different from that observed for the melting point in several systems reported. In future work, we plan to investigate the off-stoichiometric compositions that exhibit high MR and determine the effect of 3-D nanoscale confinement on the p-type or n-type materials. This will provide an interesting comparison with the results reported for silver chalcogenides thin films[7,28] and wires,[9,29] corresponding to 1-D and 2-D confinement, respectively.

This work was supported by the Office of Science, US Department of Energy, under Contract W-31-109-ENG-38, the Centre National de la Recherche Scientifique and the Université d'Orléans. We acknowledge helpful discussions with D. L. Price and thank the staff of the Advanced Photon Source for technical assistance.

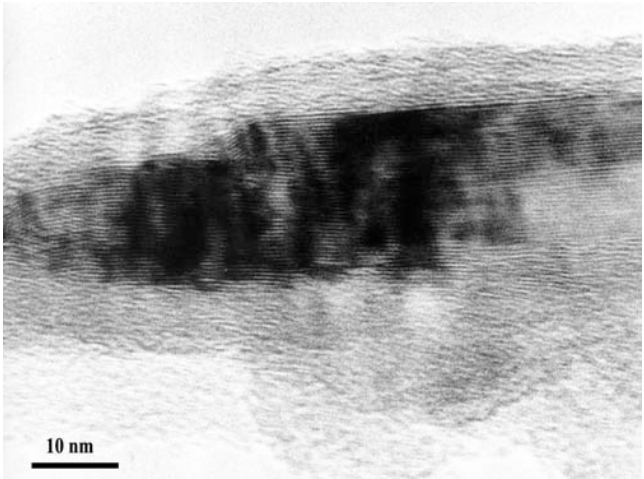

**Figure 1.** TEM picture of the silver selenide inside mesoporous silica.

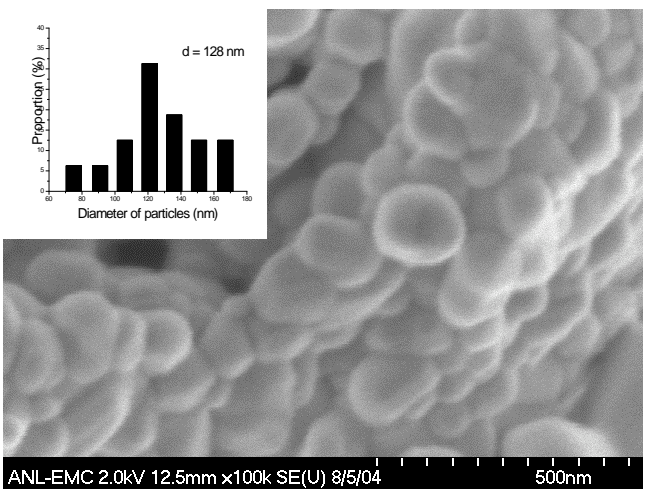

**Figure 2**. TEM picture of particles obtained by the method proposed by Dusastre et al.[21]

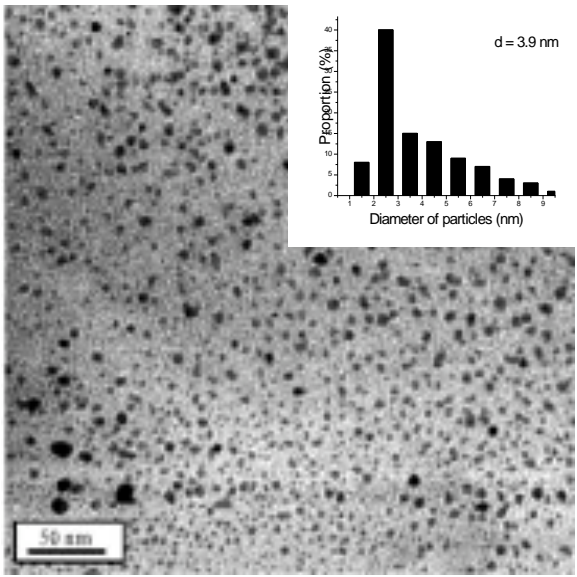

**Figure 3.** TEM pictures of nanoparticles obtained by the double microemulsion method.

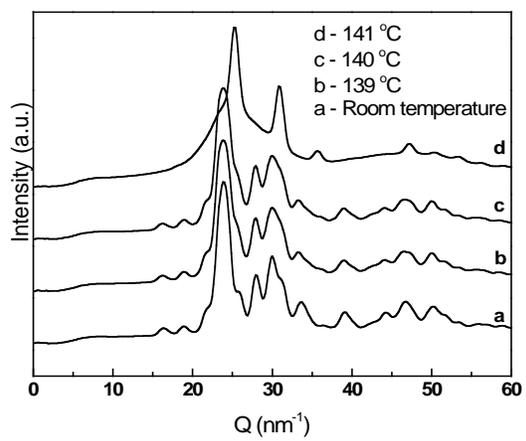

**Figure 4.** XRD of Ag$_2$Se particles with an average size of 250 nm.

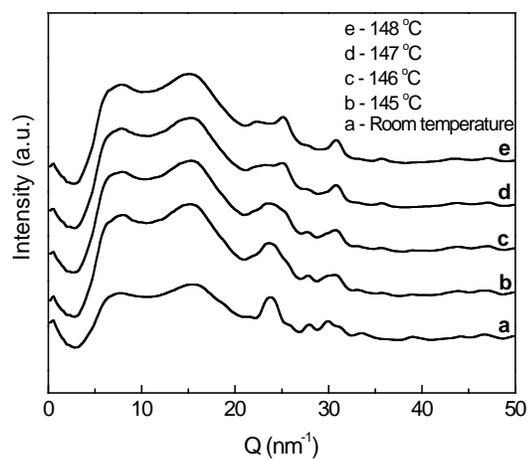

**Figure 5.** XRD of Ag$_2$Se confined inside mesoporous silica with 2 nm-sized pores.

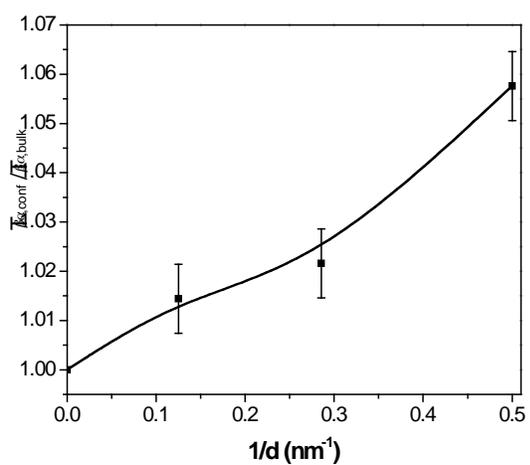

**Figure 6.** Evolution of the $\beta$-$\alpha$ phase transition temperature as a function of the silica pores diameter.

Table 1. Synthesis parameters and pore size of mesoporous silica.

| Template | Block copolymer | Aging temperature (°C)/time (h) | Mesophase | BET surface area (m$^2$/g) | Pore size (nm) |
|---|---|---|---|---|---|
| EO$_{100}$PO$_{39}$EO$_{100}$ | Pluronic F88 | 80/12 | cubic | 700 | 2.0 |
| EO$_{20}$PO$_{70}$EO$_{20}$ | Pluronic P123 | 80/12 | hexagonal | 690 | 3.5 |
| EO$_{20}$PO$_{70}$EO$_{20}$ | Pluronic P123 | 80/48 | hexagonal | 650 | 8.0 |